\RequirePackage{lineno}
\documentclass[aps,prd,superscriptaddress,twocolumn, showpacs, showkeys, amsmath, amssymb, floatfix]{revtex4} 
\usepackage[colorlinks=true, citecolor=green, filecolor=blue, linkcolor=blue, urlcolor=blue]{hyperref}
\usepackage{color}
\usepackage{graphicx}
\usepackage[perpage]{footmisc}

\newcommand \keVcc{$\mathrm{keV}/\mathrm{c}^2$}

\newcommand{\Xehund}{{XENON100}}

\newcommand{\gAe}{$g_{Ae}$}
\newcommand{\gAg}{$g_{A\gamma}$}
\newcommand{\gAN}{$g_{AN}$}

\newcommand{\columbia}{\affiliation{Physics Department, Columbia University, New York, NY, USA}}
\newcommand{\nikhef}{\affiliation{Nikhef  and the University of Amsterdam, Science Park, Amsterdam, Netherlands}}
\newcommand{\losangeles}{\affiliation{Physics \& Astronomy Department, University of California, Los Angeles, CA, USA}}
\newcommand{\assergi}{\affiliation{INFN, Laboratori Nazionali del Gran Sasso, Assergi (AQ), Italy}}
\newcommand{\coimbra}{\affiliation{Department of Physics, University of Coimbra, Coimbra, Portugal}}
\newcommand{\zurich}{\affiliation{Physics Institute, University of Z\"{u}rich, Z\"{u}rich, Switzerland}}
\newcommand{\mainz}{\affiliation{Institut f\"ur Physik \& Exzellenzcluster PRISMA, Johannes Gutenberg-Universit\"at Mainz, Mainz, Germany}}
\newcommand{\weizmann}{\affiliation{Department of Particle Physics and Astrophysics, Weizmann Institute of Science, Rehovot, Israel}}
\newcommand{\munster}{\affiliation{Institut f\"ur Kernphysik, Wilhelms-Universit\"at M\"unster, M\"unster, Germany}}
\newcommand{\purdue}{\affiliation{ Department of Physics, Purdue University, West Lafayette, IN, USA}}
\newcommand{\subatech}{\affiliation{SUBATECH, Ecole des Mines de Nantes, CNRS/In2p3, Universit\'e de Nantes, Nantes, France}}
\newcommand{\torino}{\affiliation{INFN-Torino and Osservatorio Astrofisico di Torino, Torino, Italy}}
\newcommand{\shanghai}{\affiliation{Department of Physics \& Astronomy, Shanghai Jiao Tong University, Shanghai, China}}
\newcommand{\bologna}{\affiliation{University of Bologna and INFN-Bologna, Bologna, Italy}}
\newcommand{\heidelberg}{\affiliation{Max-Planck-Institut f\"ur Kernphysik, Heidelberg, Germany}}
\newcommand{\houston}{\affiliation{Department of Physics and Astronomy, Rice University, Houston, TX, USA}}
\newcommand{\bern}{\affiliation{Albert Einstein Center for Fundamental Physics, University of Bern, Bern, Switzerland}}

\begin{document}
% \linenumbers 

\title{First Axion Results from the \Xehund\ Experiment}

\author{E.~Aprile}\columbia
\author{F.~Agostini}\thanks{Also with GSSI, INFN, L'Aquila, Italy} \bologna
\author{M.~Alfonsi}\nikhef
\author{K.~Arisaka}\losangeles
\author{F.~Arneodo} \thanks{Present address: New York University in Abu Dhabi, UAE.}\assergi
\author{M.~Auger}\zurich
\author{C.~Balan}\coimbra
\author{P.~Barrow}\zurich
\author{L.~Baudis}\zurich
\author{B.~Bauermeister}\mainz
\author{A.~Behrens}\zurich
\author{P.~Beltrame}
\thanks{Present address: SoPA, The University of Edinburgh, Edinburgh, United Kingdom.}
\email{paolo.beltrame@ed.ac.uk}\weizmann
\author{K.~Bokeloh}\munster
\author{A.~Brown}\purdue
\author{E.~Brown}\munster
\author{S.~Bruenner}\heidelberg
\author{G.~Bruno}\assergi
\author{R.~Budnik}\weizmann
\author{J.~M.~R.~Cardoso}\coimbra
\author{A.~P.~Colijn}\nikhef
\author{H.~Contreras}\columbia
\author{J.~P.~Cussonneau}\subatech
\author{M.~P.~Decowski}\nikhef
\author{E.~Duchovni}\weizmann
\author{S.~Fattori}\mainz
\author{A.~D.~Ferella}\assergi
\author{W.~Fulgione}\torino
\author{F.~Gao}\shanghai
\author{M.~Garbini}\bologna
\author{C.~Geis}\mainz
\author{L.~W.~Goetzke}\columbia
\author{C.~Grignon}\mainz
\author{E.~Gross}\weizmann
\author{W.~Hampel}\heidelberg
\author{R.~Itay}\weizmann
\author{F.~Kaether}\heidelberg
\author{G.~Kessler}\zurich
\author{A.~Kish}\zurich
\author{H.~Landsman}\weizmann
\author{R.~F.~Lang}\purdue
\author{M.~Le~Calloch}\subatech
\author{D.~Lellouch}\weizmann
\author{C.~Levy}\munster
\author{S.~Lindemann}\heidelberg
\author{M.~Lindner}\heidelberg
\author{J.~A.~M.~Lopes}\thanks{Also with Coimbra Engineering Institute, Coimbra, Portugal.}\coimbra
\author{K.~Lung}\losangeles
\author{A.~Lyashenko}\losangeles
\author{S.~MacMullin}\purdue
\author{T.~Marrod\'an~Undagoitia}\heidelberg
\author{J.~Masbou}\subatech
\author{F.~V.~Massoli}\bologna
\author{D.~Mayani~Paras}\zurich
\author{A.~J.~Melgarejo~Fernandez}\columbia
\author{Y.~Meng}\losangeles
\author{M.~Messina}\columbia
\author{B.~Miguez}\torino
\author{A.~Molinario}\torino
\author{M.~Murra}\munster
\author{J.~Naganoma}\houston
\author{K.~Ni}\shanghai
\author{U.~Oberlack}\mainz
\author{S.~E.~A.~Orrigo}\thanks{Present address: IFIC, CSIC-Universidad de Valencia, Valencia, Spain.}\coimbra
\author{E.~Pantic}\losangeles
\author{R.~Persiani}\bologna
\author{F.~Piastra}\zurich
\author{J.~Pienaar}\purdue
\author{G.~Plante}\columbia
\author{N.~Priel}\email{nadav.priel@weizmann.ac.il}\weizmann
\author{S.~Reichard}\purdue
\author{C.~Reuter}\purdue
\author{A.~Rizzo}\columbia
\author{S.~Rosendahl}\munster
\author{J.~M.~F.~dos~Santos}\coimbra
\author{G.~Sartorelli}\bologna
\author{S.~Schindler}\mainz
\author{J.~Schreiner}\heidelberg
\author{M.~Schumann}\bern
\author{L.~Scotto~Lavina}\subatech
\author{M.~Selvi}\bologna
\author{P.~Shagin}\houston
\author{H.~Simgen}\heidelberg
\author{A.~Teymourian}\losangeles
\author{D.~Thers}\subatech
\author{A.~Tiseni}\nikhef
\author{G.~Trinchero}\torino
\author{O.~Vitells}\weizmann
\author{H.~Wang}\losangeles
\author{M.~Weber}\heidelberg
\author{C.~Weinheimer.}\munster

\collaboration{The XENON100 Collaboration}\noaffiliation

%\date{\today}

\begin{abstract} 
We present the first results of searches for axions and axion-like-particles with the \Xehund\ experiment. The axion-electron coupling constant, \gAe, has been probed by exploiting the axio-electric effect in liquid xenon. A profile likelihood analysis of 224.6 live days $\times$ 34 kg exposure has shown no evidence for a signal. 
By rejecting \gAe\, larger than $7.7 \times 10^{-12}$ (90\% CL) in the solar axion search, we set the best limit to date on this coupling. In the frame of the DFSZ and KSVZ models, we exclude QCD axions heavier than 0.3 eV/c$^2$ and 80 eV/c$^2$, respectively. 
For axion-like-particles, under the assumption that they constitute the whole abundance of dark matter in our galaxy, we constrain \gAe\, to be lower than $1 \times 10^{-12}$ (90\% CL) for mass range from 1 to 40 keV/c$^2$, and set the best limit to date as well.
\end{abstract}


\maketitle 

\section{Introduction}

Axions were introduced in the Peccei-Quinn solution of the strong CP problem as pseudo-Nambu-Goldstone bosons emerging from the breaking of a global U(1) symmetry~\cite{pecceiquinn1977,weinberg1978, wilczeck1978}. Although this original model has been ruled out, ``invisible'' axions arising from a higher symmetry-breaking energy scale are still allowed, as described, for example, in the DFSZ and KSVZ models~\cite{DFS, Z, K, SVZ}. In addition to QCD axions, axion-like particles (ALPs) are pseudoscalars that do not necessarily solve the strong CP problem, but which have been introduced by many extensions of the Standard Model of particle physics. Axions as well as ALPs are well motivated cold dark matter candidates~\cite{Abbott:1982af}.

Astrophysical observations are thought to be the most sensitive technique for detecting axions and ALPs~\cite{sikivie1983}: the Sun would constitute an intense source of this particles (referred to as solar axions), where they can be produced via Bremsstrahlung, Compton scattering, axio-recombination and axio-deexcitation~\cite{redondo}. Additionally, searches can be conducted for ALPs that may have been generated via a non-thermal production mechanism in the early universe and which now constitute the dark matter in our galaxy (referred to as galactic ALPs). \\
Axions and ALPs may give rise to observable signatures in detectors through their coupling to photons (\gAg), electrons (\gAe) and nuclei (\gAN). The coupling \gAe\, may be tested via scattering off the electron of a target, such as liquid xenon (LXe), through the axio-electric effect~\cite{dimopoulos1986, avignone1987, pospelov2008, derevianko2010, arisaka2013}. This process is the analogue of the photo-electric process with the absorption of an axion instead of a photon.

We report on the first axion searches performed with the \Xehund\, experiment. The expected interaction rate is obtained by the convolution of the flux and the axio-electric cross section. The latter is given, both for QCD axions and ALPs, by
\begin{equation}
\sigma_{Ae} = \sigma_{pe}(E_A)\frac{{g_{Ae}}^2}{\beta_A}{\frac{3{E_A}^2}{16\pi\,\alpha_{em}\,{m_e}^2}}\left(1-\frac{\beta_A^{2/3}}{3} \right),
\label{eq:sigmaA} 
\end{equation}
as described in~\cite{avignone1987, pospelov2008, derevianko2010, arisaka2013, cuore2013}. In Eq.(\ref{eq:sigmaA}), $\sigma_{pe}$ is the photoelectric cross section for LXe~\cite{xpe}, $E_A$ is the axion energy, $\alpha_{em}$ is the fine structure constant, $m_e$ is the electron mass, and  $\beta_A$ is the axion velocity over the speed of light, $c$. 

The solar axion flux has recently been recalculated in~\cite{redondo}. This incorporates four production mechanisms that depend upon $g_{Ae}$: Bremsstrahlung, Compton scattering, atomic recombination, and atomic deexcitation. The corresponding flux is 30\% larger than previous estimates due to atomic recombination and deexcitation, which previously were not taken into account. However, \cite{redondo} does not include corrections for axions heavier than 1 \keVcc, which we therefore takes as an upper mass limit for our analysis. For solar axions, both flux and cross-section depend upon $g_{Ae}^2$, thus the interaction rate scales with the fourth power of the coupling.

For non-relativistic ALPs in the galaxy, assuming that they constitute the whole dark matter halo density ($\rho_{DM} \sim 0.3$ GeV/cm$^3$ \cite{Green:2011bv}), the total flux is given by $\phi_\text{ALP} = c \beta_A \times \rho_{DM} / m_A$, where $m_A$ is the ALP mass. The interaction rate for these ALPs depends on $g_{Ae}^2$, as the flux is independent from the axion coupling. As $\beta_A \approx 10^{-3}$\, in the non-relativistic regime, the velocities cancel out in the convolution between $\sigma_{Ae}$ and the flux. Thus the expected electron recoil spectrum is independent from the particle speed. As the kinetic energy of the ALPs is negligible with respect to their rest mass, a monoenergetic peak at the axion mass is expected in the spectrum. 

\section{XENON100}

The \Xehund\ experiment's primary aim is to detect dark matter in form of Weakly Interactive Massive Particle (WIMP) through their elastic nuclear scattering off nuclei in the liquid Xe target (LXe). The detector is a cylindrical (30 cm height $\times$ 30 cm diameter) dual phase time-projection chamber (TPC) with 62 kg of LXe, employed both as target and detection medium. It operates at the Laboratori Nazionali del Gran Sasso (LNGS). The detector is equipped of 242 radio pure photomultiplier tubes (PMTs) placed on top (in the xenon gas) and on the bottom of the TPC (immersed in the LXe below the cathode). A particle interaction in the LXe target creates both excited and ionized atoms. De-excitation leads to a prompt scintillation signal ($S1$). Due to the presence of an electric drift field of 530 V/cm, a large fraction of the ionization electrons is drifted away from the interaction site and extracted from the liquid into the gas phase by a strong extraction field of $\sim$12 kV/cm, generating a light signal ($S2$) by proportional scintillation in the gas. Three-dimensional event vertex reconstruction is achieved using the time difference between the $S1$ and the $S2$ signals along with the $S2$-hit-pattern on the top PMTs, which is employed to estimate the (x\,,\,y)\,-\,coordinate. The $S1$ signal is used to estimate the energy deposited in the detector, as explained below (Eq.~\ref{eq:Ly}). A detailed description of the instrument is given in~\cite{xe100_instr2012}.

The ratio $S2/S1$ is different whether the energy deposit in the LXe is due to electronic recoil (ER) or to nuclear recoil (NR). Therefore, this $S2/S1$ ratio is used to discriminate the two topologies of events. In the case of ERs, such as from interaction with $\gamma$, $\beta$ backgrounds and axion signals, the energy from the incoming particle is transferred to the electrons of the Xe atom. Conversely, neutrons or WIMPs scatter off the Xe nuclei. \\
The total background in the inner 34 kg super-ellipsoidal fiducial volume of the LXe target corresponds to $\mathrm{5.3 \times 10^{-3}\, events/(keV \times kg \times day)}$~\cite{xe100_si2012}, making \Xehund\ extremely sensitive to rare event searches in general. The ultra low background has been achieved by means of several techniques: the careful selection of materials~\cite{xe100_screening}; the detector design, with radioactive parts far away from the target; the powerful passive shield as well as an active LXe veto; the self-shielding power of LXe, exploited by selecting only the inner part of the TPC for the analysis. The background is dominated by Compton events which scatter only once in the low-energetic region of interest, resulting in an almost flat spectrum~\cite{xe100_erbkg2011}.
Under an average depth of 3600\,m water equivalent, the cosmic muon flux is suppressed by six orders of magnitude with respect to sea level.

\section{Analysis}

\subsection{Data sample and analysis}

In this work, we analyse the same data set used for the spin-independent~\cite{xe100_si2012} and spin-dependent~\cite{xe100_sd2013} WIMP-searches, with an exposure of 224.6 live days and 34 kg fiducial mass. Two main classes of analysis cuts have been applied. The first one consists of basic data quality selection, to remove either unidentified energy deposition peaks or excessive electronic noise level. Since only single-scatter events are expected from axion interactions, the second class of cuts identifies such events by using the number of $S1$ and $S2$ peaks. Conditions on the size of the $S2$ and the requirement that at least two PMTs must observe an $S1$ signal ensure that only data above the threshold and well above the noise level are considered. Finally, consistency criteria are applied. These are identical to the one used in the above mentioned WIMP searches, with the exception of a cut on the $S2$ width. The original definition of this consistency cut, comparing the width of the proportional $S2$ 
peak to its time delay with respect to the $S1$, had been found to not be useful for this analysis targeted at ERs, and was hence not used. Detailed information on the procedure is available in~\cite{xe100_ana2012}.

%%%%%%%%%
\begin{figure}[t!]
\begin{minipage}{1.\linewidth}
\centerline{\includegraphics[width=1.\linewidth]{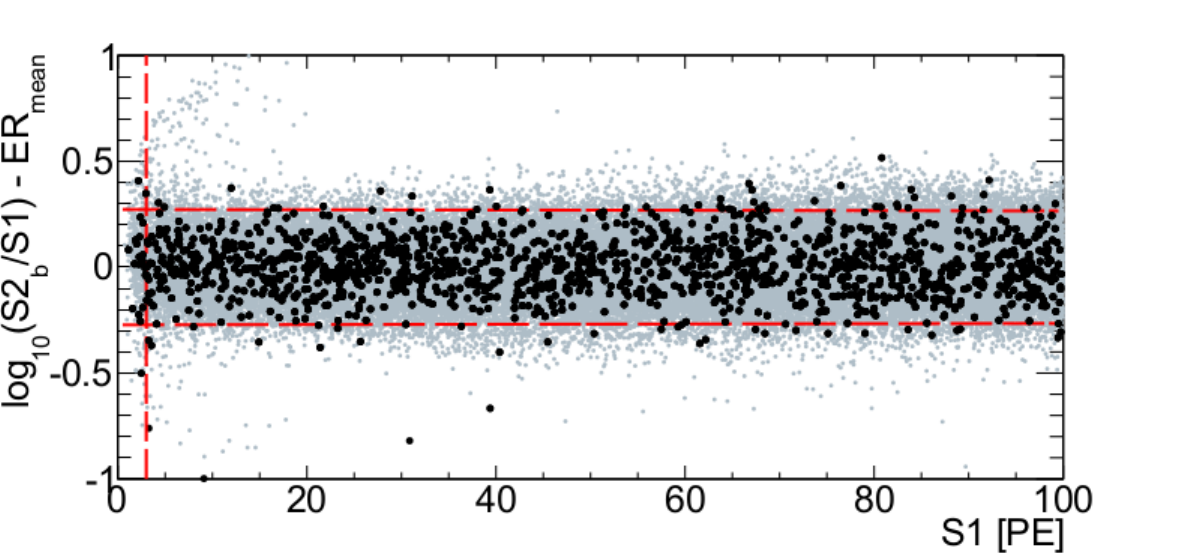}}
\centerline{\includegraphics[width=1.\linewidth]{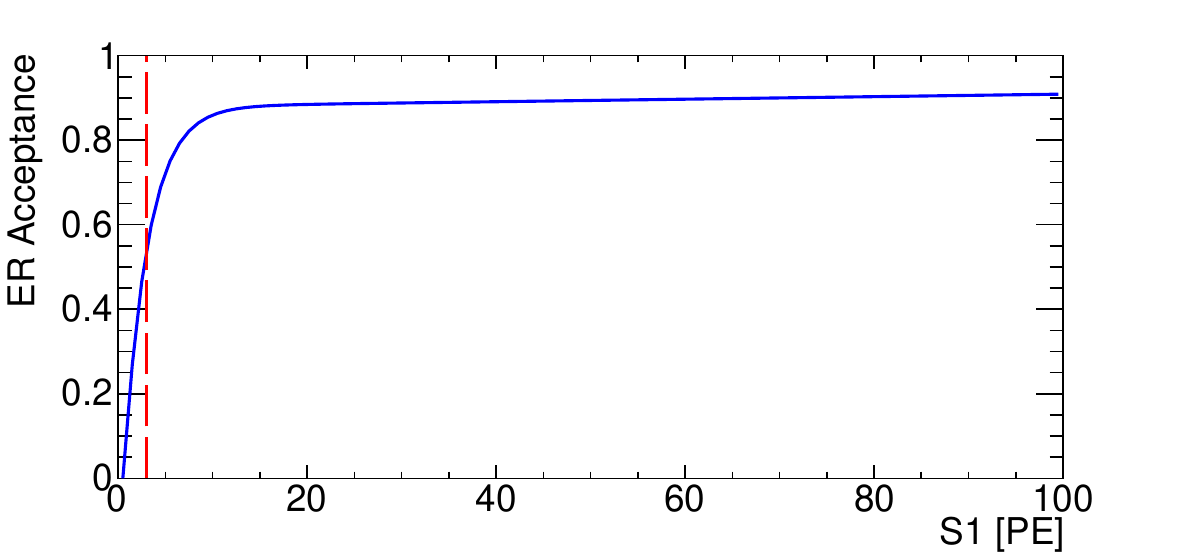}}
\end{minipage}
\caption{{\it Top}: Event distribution in the flattened $\mathrm{log_{10}}(S2_b/S1)$ vs. $S1$ space for science data (black points) and calibration (grey points). Straight dashed lines show the selection cut on the flattened $\mathrm{log_{10}}(S2_b/S1)$ (horizontal red lines) and the 3 PE threshold cut (red vertical line). {\it Bottom}: Global acceptance for electronic recoil events, evaluated on calibration data.}
\label{fig:dm_band_eff}
\end{figure}
%%%%%%%%% 
Figure~\ref{fig:dm_band_eff} (top) shows the distribution in the $\mathrm{log_{10}}(S2_b/S1)$ vs.$\, S1$ for calibration data (grey dots), and the science data passing all the selection cuts (black dots). Only the $S2$ signal detected by the bottom PMTs, $S2_b$, is used since it requires smaller corrections~\cite{xe100_instr2012}. The calibration data is obtained by exposing the detector to $^{60}$Co and $^{232}$Th sources. These have been chosen as their high energy gamma rays can penetrate the LXe into the fiducial volume, leaving a low-energetic Compton scatter spectrum covering the energy region of interest. The mean of the $\mathrm{log_{10}}(S2_b/S1)$ band from the calibration is subtracted in order to remove the energy-dependence of this parameter. The lower energy threshold was set to 3 photoelectrons (PE) in $S1$ ($\sim 2$ keV for ER energy deposit) in order to limit the presence of random coincidences from dark counts in the PMTs. In addition, a lower threshold of 150 PE in $S2$ has been imposed to be 
unaffected by the trigger threshold~\cite{xe100_ana2012}.

In order to reject ER events with an anomalously high or low $S2/S1$ ratio, signal candidates are required to be inside the $2 \sigma$ band around the $\mathrm{log_{10}}(S2_b/S1)$ median~\cite{xe100_ana2012}. This is shown by the horizontal red dashed lines in Fig.~\ref{fig:dm_band_eff} (top). The combined acceptance of all selection cuts for ER events is evaluated on calibration data, and is shown in Fig.~\ref{fig:dm_band_eff} (bottom). Upper thresholds of 30 and 100 PE were employed for the axions from the Sun and the non-relativistic ALPs searches, respectively.

%%%%%%%%%
\begin{figure}[t!]
\begin{minipage}{1.\linewidth}
\centerline{\includegraphics[width=1.\linewidth]{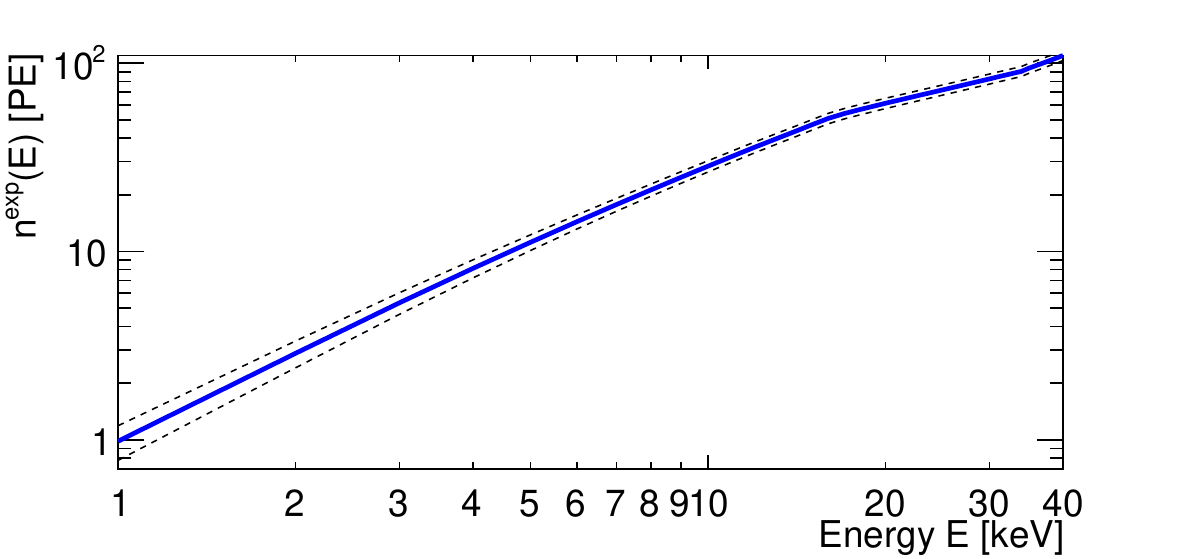}}
\end{minipage}
\caption{Conversion function between energy recoil in keV and $S1$ in PE. The $n^{exp}$ central value and the $\pm 1 \sigma$ uncertainty are indicated with solid blue and black dashed line, respectively.}
\label{fig:ly}
\end{figure}
%%%%%%%%% 
The energy deposited by each interaction is obtained using the observed $S1$ signal. The keV - PE conversion is performed using the NEST model (v0.98)~\cite{Szydagis:2011tk}. This takes into account the scintillation efficiency $R(E)$ relative to the 32.1 keV transition of $^{83m}$Kr at zero electric field (as chosen by~\cite{aprile_compton} and~\cite{baudis_compton}) and the quenching factor $Q(E)$ for a non-zero electric field (measured by~\cite{baudis_compton} for values close to the field applied in XENON100). The model agrees with the direct measurements at zero field~\cite{aprile_compton,baudis_compton}, as well as the measurements with a non-zero field~\cite{manalaysay_field,baudis_compton}. The uncertainty on $R(E) \times Q(E)$ is taken from NEST and assumed to be Gaussian. This reflects the intrinsic uncertainty of the model (4\%) as well as the spread in the measured data points, particularly relevant at low energies. The conversion from the energy deposition $E$ to the observed signal $n^{exp}$ in PE is therefore given by
\begin{equation}
n^{exp}(E) = R(E) \times Q(E) \times f \times E \equiv L_Y(E) \times E,
\label{eq:Ly}
\end{equation}
where the factor $f = 3.76$ PE/keV is the derived \Xehund\ light yield at 32.1\,keV and zero field ~\cite{xe100_instr2012, manalaysay_field}. The function $n^{exp}(E)$ is shown in Fig.~\ref{fig:ly}, together with the $\pm 1 \sigma$ uncertainty.

\subsection{Statistical method}

A Profile Likelihood analysis, as described in~\cite{asympl} and analogous to~\cite{xe100_pl}, is used to constrain the coupling constant \gAe. The full likelihood function is given by 
\begin{equation}
\mathcal{L} = \mathcal{L}_1(g_{Ae}, N_b, n^{exp}) \times \mathcal{L}_2(n^{exp}),
\label{eq:likelihood}
\end{equation}
The parameter of interest is \gAe, whereas $N_b$ and $n^{exp}$ are considered as nuisance parameters.
The first term, 
\begin{equation}
\mathcal{L}_1 = \mathrm{Poiss}(N | N_s + N_b) \prod_{i=1}^{N} \frac{N_s f_s(S1_i) + N_b f_b(S1_i)}{N_s + N_b},
\label{eq:signallike}
\end{equation}
describes the measurement of the detector. The second term,
\begin{equation}
\mathcal{L}_2(n^{exp}(t)) = \mathrm{e}^{-t^2 / 2},
\label{eq:enelike}
\end{equation}
is used to constrain the energy scale.

The energy scale term, $\mathcal{L}_2$, has been parametrised with a single parameter $t$. The likelihood function is defined to be normally distributed with zero mean and unit variance, corresponding to where $t=\pm 1$ corresponds to a $\pm$1$\sigma$ deviation in $n^{exp}$, as shown in Fig.~\ref{fig:ly}, \textit{i.e.}, $t=(n^{exp}-n^{exp}_{mean})/\sigma$.

In Eq.(\ref{eq:signallike}) $N_s$ and $N_b$ are the expected number of signal and background events in the search region, and $N_s$ depends upon $g_{Ae}$ and $n^{exp}$. $N$ is the total number of observed events, and the $S1_i$ corresponds to the $S1$ of the $i$-th event. The functions $f_s$ and $f_b$ are the normalised signal and background probability distribution functions.

The event rate with a given number of detected photons, $n$, is obtained by applying Poisson smearing to the predicted energy spectrum $dR/dE$,
\begin{equation}
 \frac { dR }{ dn } =\int _{ 0 }^{ \infty  }{ \frac { dR }{ dE  }  } \times \mathrm{Poiss}\left( n | { n }^{ exp }(E) \right) dE,
\label{eq:dRdn}
\end{equation}
where $n^{exp}$ is obtained from Eq.(\ref{eq:Ly}).

The rate as a function of the measured number of photoelectrons, $S1$, is given by
\begin{equation}
 \frac {dR}{dS1} = \sum _{ n = 1 }^{ \infty  }{ \mathrm{Gauss}(S1|n,\sqrt { n } { \sigma  }_{ PMT }) } \times \frac { dR }{ dn } \times { \epsilon  }(S1),
\label{eq:dRdS1}
\end{equation}
where ${ \sigma  }_{ PMT }=0.5$ PE is the PMT resolution~\cite{xe100_ana2012}, and ${\epsilon}(S1)$ is the acceptance of all criteria applied to the data, see Fig.~\ref{fig:dm_band_eff} (bottom). It has a rather flat behavior above 10 PE. Below that, the acceptance decreases mainly due to data quality criteria.

%%%%%%%%%
\begin{figure}[t!]
\begin{minipage}{1.\linewidth}
\centerline{\includegraphics[width=1.\linewidth]{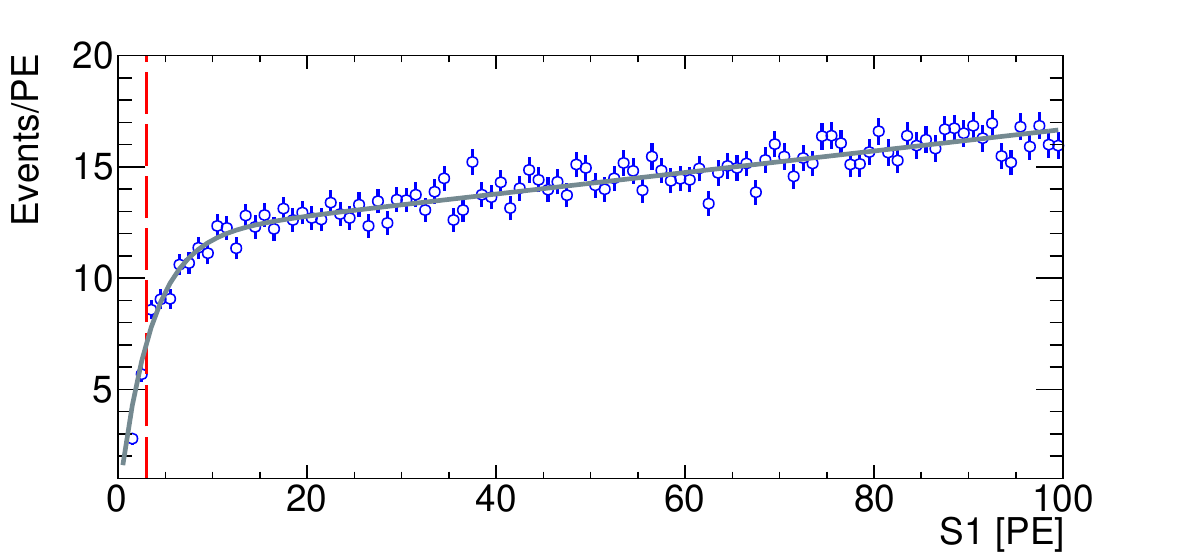}}
\end{minipage}
\caption{Background model $N_b \times f_b$ (grey line), scaled to the correct exposure, as explained in the text. The analytic function $f_b$ is based on the $^{60}$Co and $^{232}$Th calibration data (empty blue dots), and is used in Eq.(\ref{eq:signallike}). The 3 PE threshold is indicated by the vertical red dashed line.}
\label{fig:bkgmodel}
\end{figure}
%%%%%%%%%
The majority of the background events arises from gamma scattering off the atomic electrons of the LXe target, as well from intrinsic beta-background ($^{222}$Rn and $^{85}$Kr)~\cite{xe100_erbkg2011}. To model these events, we use the $^{60}$Co and $^{232}$Th calibration data. The total spectrum is then analytically parametrised by means of a modified Fermi function, $f_b(S1)$, shown in Fig.~\ref{fig:bkgmodel} (grey line) along with the calibration data (empty blue dots). The spectrum is scaled to the science data exposure by normalizing it to the number of events seen outside the signal region, to avoid biases. For solar axions, it is done between 30 and 100 PE, and for galactic ALPs below $m_{A}[$pe$]-2\sigma$ and above $m_{A}[$pe$]+2\sigma$, where $m_{A}[$pe$]$ is the ALP mass in units of PE and $\sigma$ is the width of the expected signal peak, see Fig.~\ref{fig:galactic_s1rate}. Then, the scaled background spectrum is integrated in the signal region to give the expected number of background events, $N_
b$. The background model scaled to the correct exposure, $N_b~\times~f_b$, is shown in Fig.~\ref{fig:bkgmodel}, along with the 
scaled calibration spectrum.

As downward statistical fluctuations of the background might lead to reject couplings to which the experiment is not sensitive, we used the $\mathrm{CL_s}$ method to protect the result from this effect, as described in~\cite{xe100_pl}.

\section{Results}

\subsection{Solar axions}

\begin{figure}[t!]
\begin{minipage}{1.\linewidth}
\centerline{\includegraphics[width=1.\linewidth]{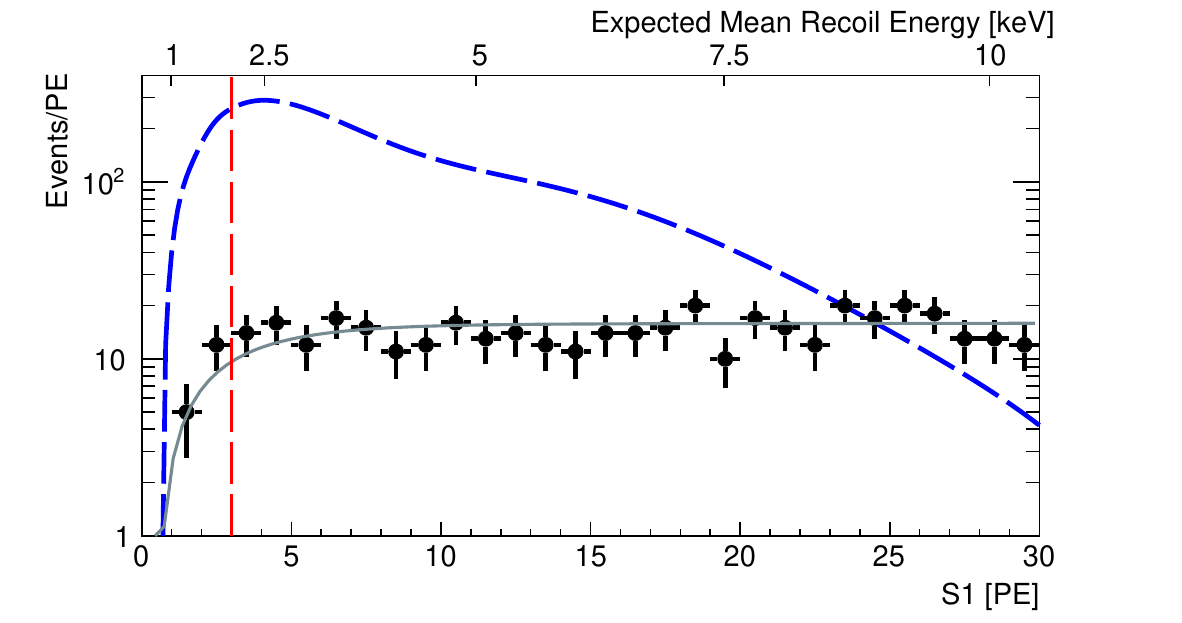}}
\end{minipage}
\caption{Event distribution of the data (black dots), and background model (grey) of the solar axion search. The expected signal for solar axions with  $m_A<$ 1 keV/c$^2$ is shown by the dashed blue line, assuming \gAe $ = 2 \times 10^{-11}$, the current best limit, from EDELWEISS-II~\cite{edelweiss2013}. The vertical dashed red line indicates the low $S1$ threshold, set at 3 PE. The top axis indicates the expected mean energy for electronic recoils as derived from the observed S1 signal.}
\label{fig:solar_s1rate}
\end{figure}

The spectrum of the remaining 393 events, between 3 and 30 PE and after all the selection cuts, are shown in Fig.~\ref{fig:solar_s1rate} as a function of $S1$. The solid grey line shows the background model, $N_b \times f_b$. The expected $S1$ spectrum for solar axions, lighter than 1 \keVcc, is shown as a blue dashed line for \gAe $ = 2 \times 10^{-11}$, {\it i.e.} the best limit so far, reported by the EDELWEISS-II collaboration~\cite{edelweiss2013}. The data are compatible with the background model, and no excess is observed for the background only hypothesis.

%%%%%%%%%
%%%%%%%%%
\begin{figure}[h!]
\begin{minipage}{1.\linewidth}
\centerline{\includegraphics[width=1.\linewidth]{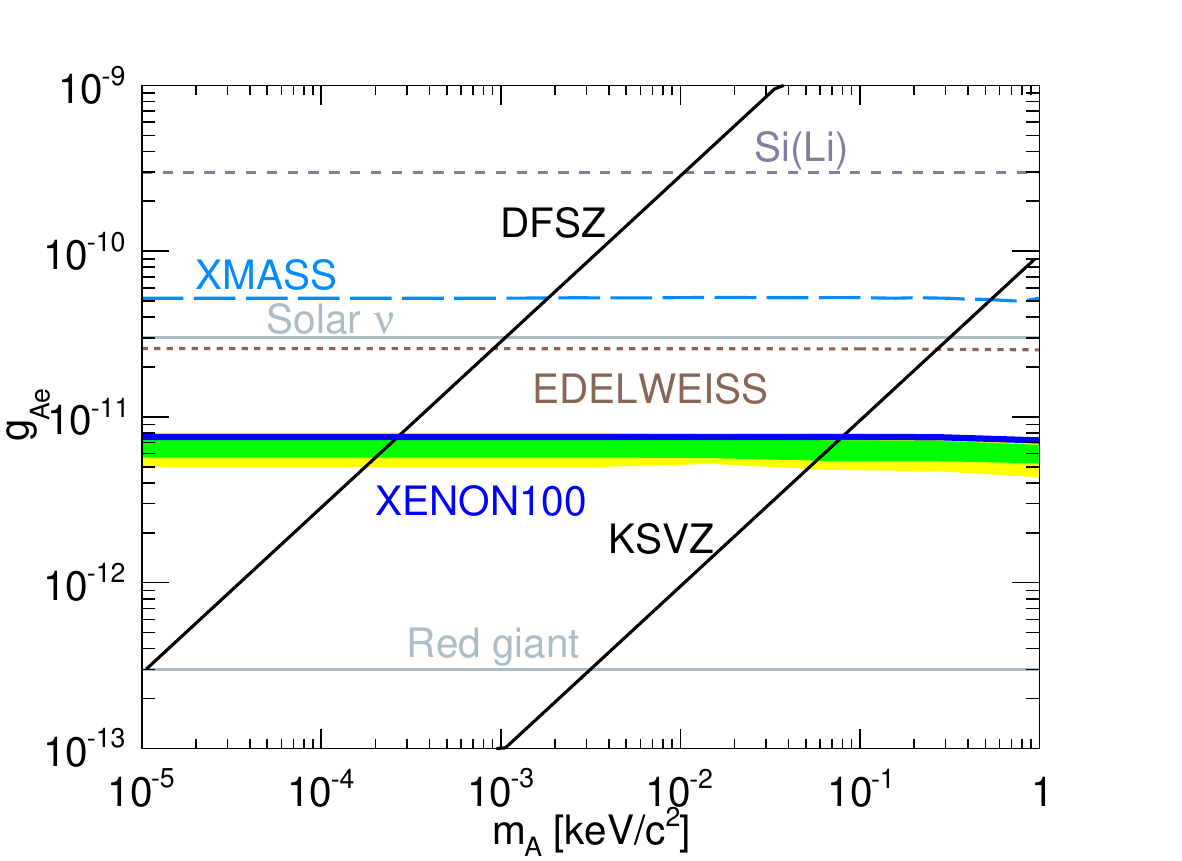}}
\end{minipage}
\caption{The \Xehund\ limits (90\% CL) on solar axions is indicated by the blue line. The expected sensitivity, based on the background hypothesis, is shown by the green/yellow bands ($1 \sigma$/$2 \sigma$). 
Limits by EDELWEISS-II~\cite{edelweiss2013}, and XMASS~\cite{xmass2013} are shown, together with the limits from a Si(Li) detector from Derbin et al.~\cite{derbin2012}. Indirect astrophysical bounds from solar neutrinos~\cite{gondolo2009} and red giants~\cite{viaux2013} are represented by dashed lines. The benchmark DFSZ and KSVZ models are represented by black lines~\cite{DFS, Z, K, SVZ}.}
\label{fig:gAe_Solar_sensitivity-exclusion}
\end{figure}
%%%%%%%%%

Figure~\ref{fig:gAe_Solar_sensitivity-exclusion} shows the new \Xehund\ exclusion limit on \gAe\, at 90\% CL. The sensitivity is shown by the green/yellow band ($1 \sigma / 2 \sigma$). As we used the most recent and accurate calculation for solar axion flux from~\cite{redondo}, which is valid only for light axions, we restrict the search to $m_A < 1$ \keVcc. For comparison, we also present other recent experimental constraints~\cite{derbin2012, xmass2013, edelweiss2013}. Astrophysical bounds~\cite{gondolo2009, raffelt2008, viaux2013} and theoretical benchmark models~\cite{DFS, Z, K, SVZ} are also shown. For solar axions with masses below 1 \keVcc\ \Xehund\, is able to set the strongest constraint on the coupling to electrons, excluding values of \gAe \,larger than $7.7 \times 10^{-12}$ (90\% CL).

For a specific axion model the limit on the dimensionless coupling \gAe\ can be translated to a limit on the axion mass. Within the DFSZ and KSVZ models~\cite{DFS, Z, K, SVZ} \Xehund\ excludes axion masses above 0.3 eV/c$^2$ and 80 eV/c$^2$, respectively. For comparison, the CAST experiment, testing the coupling to photons, \gAg, has excluded axions within the KSVZ model in the mass range between 0.64 - 1.17 eV/c$^2$~\cite{Arik2014, Barth:2013sma}.

%%%%%%%%%
\begin{figure}[t!]
\begin{minipage}{1.\linewidth}
\centerline{\includegraphics[width=1.\linewidth]{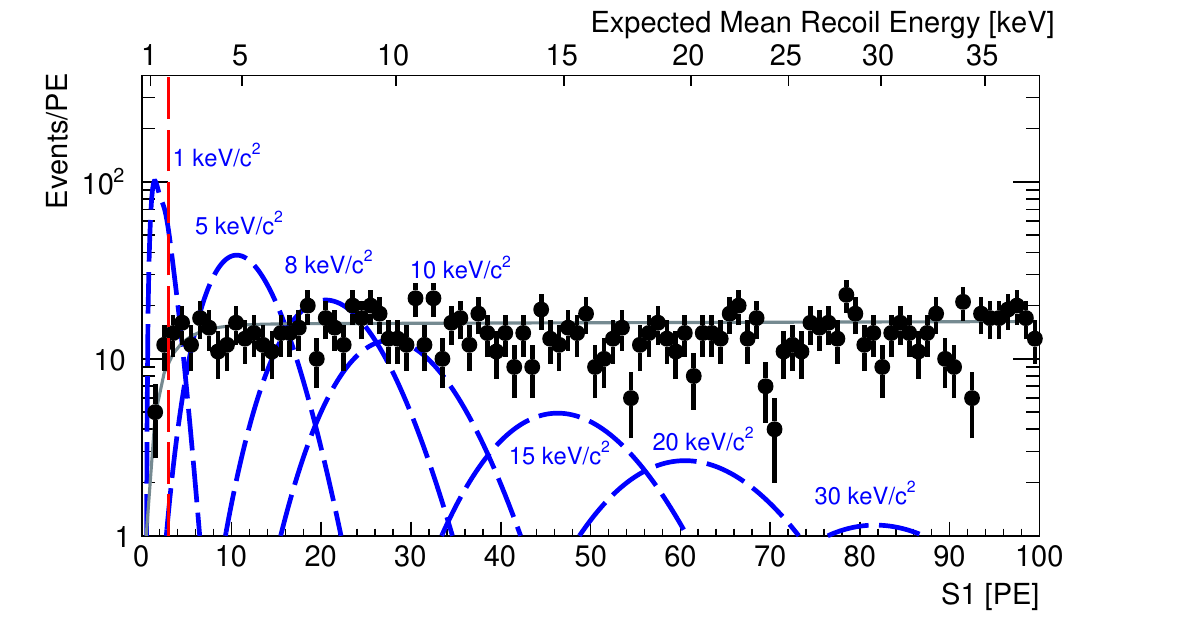}}
\end{minipage}
\caption{Event distribution in the galactic ALPs search region between 3 and 100 PE (black dots). The grey line shows the background model used for the profile likelihood function. The red dashed line indicates the $S1$ threshold. The expected signal in \Xehund\ for various ALP masses, assuming \gAe $ = 9 \times 10^{-13}$, is shown as blue dashed lines. The top axis indicates the expected mean energy for electronic recoils as derived from the observed $S1$ signal.}
\label{fig:galactic_s1rate}
\end{figure}
%%%%%%%%%
\begin{figure}[h!]
\begin{minipage}{1.\linewidth}
\centerline{\includegraphics[width=1.\linewidth]{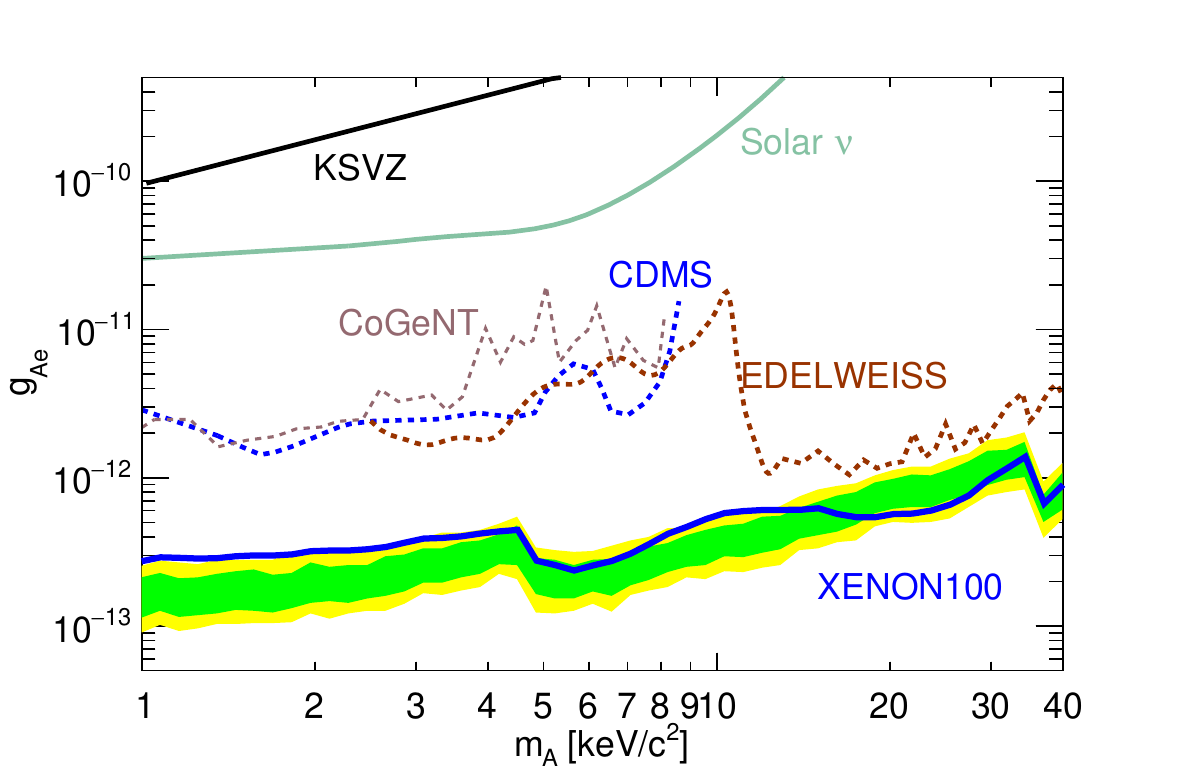}}
\end{minipage}
\caption{The \Xehund\ limit (90\% CL) on ALP coupling to electrons as a function of the mass, under the assumption that ALPs constitute all the dark matter in our galaxy (blue line). The expected sensitivity is shown by the green/yellow bands ($1 \sigma / 2 \sigma$). The other curves are constraints set by CoGeNT~\cite{cogent2008} (light brown dashed line), CDMS~\cite{cdms2009} (blue dashed line), and EDELWEISS-II~\cite{edelweiss2013} (brown dashed line). The indirect astrophysical bound from solar neutrinos~\cite{gondolo2009} is represented as a grey line. The benchmark KSVZ model is represented by a black line~\cite{K, SVZ}.}
\label{fig:gAe_Galactic_sensitivity-exclusion}
\end{figure}
%%%%%%%%%

\subsection{Galactic axions-like particles}

Figure~\ref{fig:galactic_s1rate} shows the  \Xehund\ data after the selection cuts in the larger energy region of interest used for the search for non-relativistic galactic ALPs (1422 surviving events), along with their statistical errors. Also shown is the expected signal for different ALP masses, assuming a coupling of \gAe $ = 9 \times 10^{-13}$ and that ALPs constitute all of the galactic dark matter. The width of the monoenergetic signal is given by the energy resolution of the detector at the relevant $S1$ signal size~\cite{xe100_instr2012}. As for the solar axion search, the data is compatible with the background hypothesis, and no excess is observed for the background-only hypothesis for the various ALP masses.

The \Xehund\, 90\% CL exclusion limit for galactic ALPs is shown in Fig.~\ref{fig:gAe_Galactic_sensitivity-exclusion}, together with other experimental constraints~\cite{cogent2008, cdms2009, edelweiss2013}. Astrophysical bounds~\cite{gondolo2009, raffelt2008, viaux2013} and the KSVZ benchmark model~\cite{K, SVZ} are also presented. The expected sensitivity is shown by the green/yellow bands ($1\sigma / 2 \sigma$). The steps in the sensitivity around 5 and 35 \keVcc\ reflect the photoelectric cross section due to the atomic energy levels.
Below 5~\keVcc\ the obtained 90\% CL is higher than expected, deviating by as much as $2 \sigma$ from the mean predicted sensitivity. This is due to a slight excess of events between 3 and 5 PE. A similar effect is responsible for the limit oscillating around the predicted sensitivity above 5~\keVcc. The ALP limit is very sensitive to fluctuations in individual bins because of the expected monoenergetic signal. In the (1\,-\,40) keV/c$^2$ mass range, \Xehund\ sets the best upper limit, excluding an axion-electron coupling $g_{Ae}>1 \times 10^{-12}$ at the 90\% CL, assuming that ALPs constitute all of the galactic dark matter.

% \subsection{Systematic uncertainties estimation}
\vskip 10pt

The impact of systematic uncertainties has been evaluated for both analyses presented here. In particular, we have considered the parametrisation of the cross section of the axio-electric effect, the data selection based on a band in the $\mathrm{log_{10}}(S2_b/S1)$ vs. $S1$ space, the choice of the fiducial volume, as well as the conversion of the $S1$ signal into an ER enegry and the energy resolution.

Previous works (e.g.~\cite{xmass2013, arisaka2013}) have used a different parametrisation of the axion velocity term in $\sigma_A$, while we chose to employ $(1-\beta_A^{2/3}/3)$, Eq.(\ref{eq:sigmaA}), as suggested by~\cite{edelweiss2013}. However, we also tested the other assumptions and found the impact on the final limit to be negligible.

Varying the width of the band chosen to select the data entering the analysis (shown in Fig.~\ref{fig:dm_band_eff} (top) as horizontal dashed red lines) from $\pm 1 \sigma$ up to $\pm 4 \sigma$ changes the final result on \gAe\, by 5\%, {\it i.e.} well within the $\pm 2 \sigma$ of the sensitivity band.

Similarly, a variation of the fiducial volume has a negligible impact on the sensitivity: the inner ellipsoid was changed in size to accomodate between 28 kg and 40 kg, but maintaining the same 224.6 days of live time. The reduced background for smaller fiducial masses is compensated by the smaller total exposure, resulting in a variation of the limit well below 10\%.

The uncertainty on the energy scale used for the conversion from the observed $S1$ signal in PE into keV, Fig.~\ref{fig:ly} and Eq.(\ref{eq:Ly}), is taken into account in the profile likelihood function and is profiled out via the nuisance parameter $t$, Eq.(\ref{eq:enelike}). The detector's energy resolution is considered by smearing the predicted energy spectrum $dR/dE$ by Poisson and Gaussian processes, as described in Eq.(\ref{eq:dRdS1}). We note that the final results on \gAe\, are also robust against further changes in the energy scale: even if $L_Y(E)$, as defined in Eq.(\ref{eq:Ly}), is varied by 25\%, the limits change by less than 5\% and about 10\% for the solar and for the galactic axion searches, respectively. 

\section{Acknowledgments}

We gratefully acknowledge support from NSF, DOE, SNF, Volkswagen Foundation, FCT, Region des Pays de la Loire, STCSM, NSFC, DFG, MPG, Stichting voor Fundamenteel Onderzoek der Materie (FOM), the Weizmann Institute of Science, the EMG research center and INFN. We are grateful to LNGS for hosting and supporting XENON100.

%%% BIBLIOGRAPHY %%%

\bibliographystyle{apsrev}
\bibliography{Xe100AxionRun10}

\end{document}